\pdfoutput=1

\documentclass[11pt]{article}

\usepackage[preprint]{acl}

\usepackage{times}
\usepackage{tabularx} 
\usepackage{latexsym}
\usepackage[normalem]{ulem} 
\usepackage{amssymb} 
\usepackage{booktabs}  
\usepackage{graphicx}
\usepackage{amsmath}
\usepackage{subcaption}
\usepackage{colortbl}  

\definecolor{lightblue}{HTML}{D9EAD3}
\definecolor{lightgreen}{HTML}{F9CB9C}
\definecolor{lightyellow}{HTML}{FFF2CC}
\definecolor{lightorange}{HTML}{FCE5CD}
\definecolor{genrecolor}{RGB}{19, 195, 235}      
\definecolor{layacolor}{RGB}{244, 143, 177}       
\definecolor{ragacolor}{RGB}{209, 19, 41}         
\definecolor{instrumentcolor}{RGB}{171, 164, 36}   

\usepackage[T1]{fontenc}

\usepackage[utf8]{inputenc}

\usepackage{microtype}

\usepackage{inconsolata}

\usepackage{graphicx}
\usepackage{hyperref}
\title{Music for All: Representational Bias and Cross-Cultural Adaptability of Music Generation Models}

\author{
    \textbf{Atharva Mehta} \quad \textbf{Shivam Chauhan} \quad \textbf{Amirbek Djanibekov} \\
    \textbf{Atharva Kulkarni} \quad \textbf{Gus Xia} \quad \textbf{Monojit Choudhury} \\
    Mohamed bin Zayed University of Artificial Intelligence\\
    \texttt{\{atharva.mehta, shivam.chauhan, amirbek.djanibekov,} \\  
    \texttt{atharva.kulkarni, gus.xia, monojit.choudhury\}@mbzuai.ac.ae}
}

\begin{document}
\maketitle
\begin{abstract}
The advent of Music-Language Models has greatly enhanced the automatic music generation capability of AI systems, but they are also limited in their coverage of the musical genres and cultures of the world. We present a study of the datasets and research papers for music generation and quantify the bias and under-representation of genres. We find that only 5.7\% of the total hours of existing music datasets come from non-Western genres, which naturally leads to disparate performance of the models across genres. 
We then investigate the efficacy of Parameter-Efficient Fine-Tuning (PEFT) techniques in mitigating this bias. Our experiments with two popular models -- MusicGen and Mustango, for two underrepresented non-Western music traditions --  Hindustani Classical and Turkish Makam music, highlight the promises as well as the non-triviality of cross-genre adaptation of music through small datasets, implying the need for more equitable baseline music-language models that are designed for cross-cultural transfer learning. The code for the paper is available at our \href{https://github.com/atharva20038/music4all}{Github Repository} and the model adapters are available at \href{https://huggingface.co/collections/athi180202/music4all-67a0778b5b562859c2a9a8e1}{Huggingface}. 

\end{abstract}

\section{Introduction}

Music, as a powerful expression of cultural identity, is deeply embedded in traditions~\cite{10.1093/mq/79.2.281, chung2006digital}. Recent advancements in AI, powered by deep learning models~\cite{c:24,c:23,c:22}, have led to significant improvements in automatic music generation technologies. This progress has led to several music generation playgrounds such as Jukebox~\cite{c:25}, Suno\footnote{https://suno.com/}, and Udio\footnote{https://www.udio.com/} offering users the ability to generate music according to their specifications. However, these models often reflect biases, particularly towards Western musical traditions~\cite{10.1093/pnasnexus/pgae346, c:23}, in their training data. 

This lack of diversity in datasets, as outlined by~\citet{c:23, melechovsky-etal-2024-mustango, c:25}, is also evident in the disparate performance of the music generation models across genres. More specifically, the models tend to rely on Western tonal and rhythmic structures when generating music for non-Western genres, such as Indian or Middle Eastern music. The situation is comparable to the lack of cultural and linguistic diversity~\cite{c:27, bender-2018, bender2021dangers} in NLP research.


In order to quantify the severity of this problem in music generation research landscape, we conduct a comprehensive analysis of existing music datasets and music generation papers, which reveals a stark disparity in the representation of non-Western music. Particularly noteworthy is the scarcity of non-Western music data, with merely 5.7\% of the total hours of the available datasets. This finding highlights the need for more diverse musical datasets and methods to adapt state-of-the-art models to low-resource genres. 

However, it remains unclear whether cross-genre music adaptation, similar to cross-lingual adaptation, can be effectively achieved using lightweight computational techniques such as parameter-efficient fine-tuning (PEFT) \cite{houlsby2019parameter}. In this paper, we explore this question by adapting two open-source models, MusicGen~\cite{c:23} and Mustango~\cite{melechovsky-etal-2024-mustango} for two low-resource non-Western genres - \textit{Hindustani Classical}~\footnote{Hindustani Classical music is a traditional system of music that emphasizes melodic development based on ragas (melodic frameworks) and talas (rhythmic cycles).} music of India and \textit{Makamat}~\footnote{Makam, in traditional Arabic music, is a melodic mode system defining pitches, patterns, and improvisation, central to Arabian art music, with 72 heptatonic scales} music of the Middle East.



\begin{figure*}[!ht]
\centering
\includegraphics[width=\textwidth]{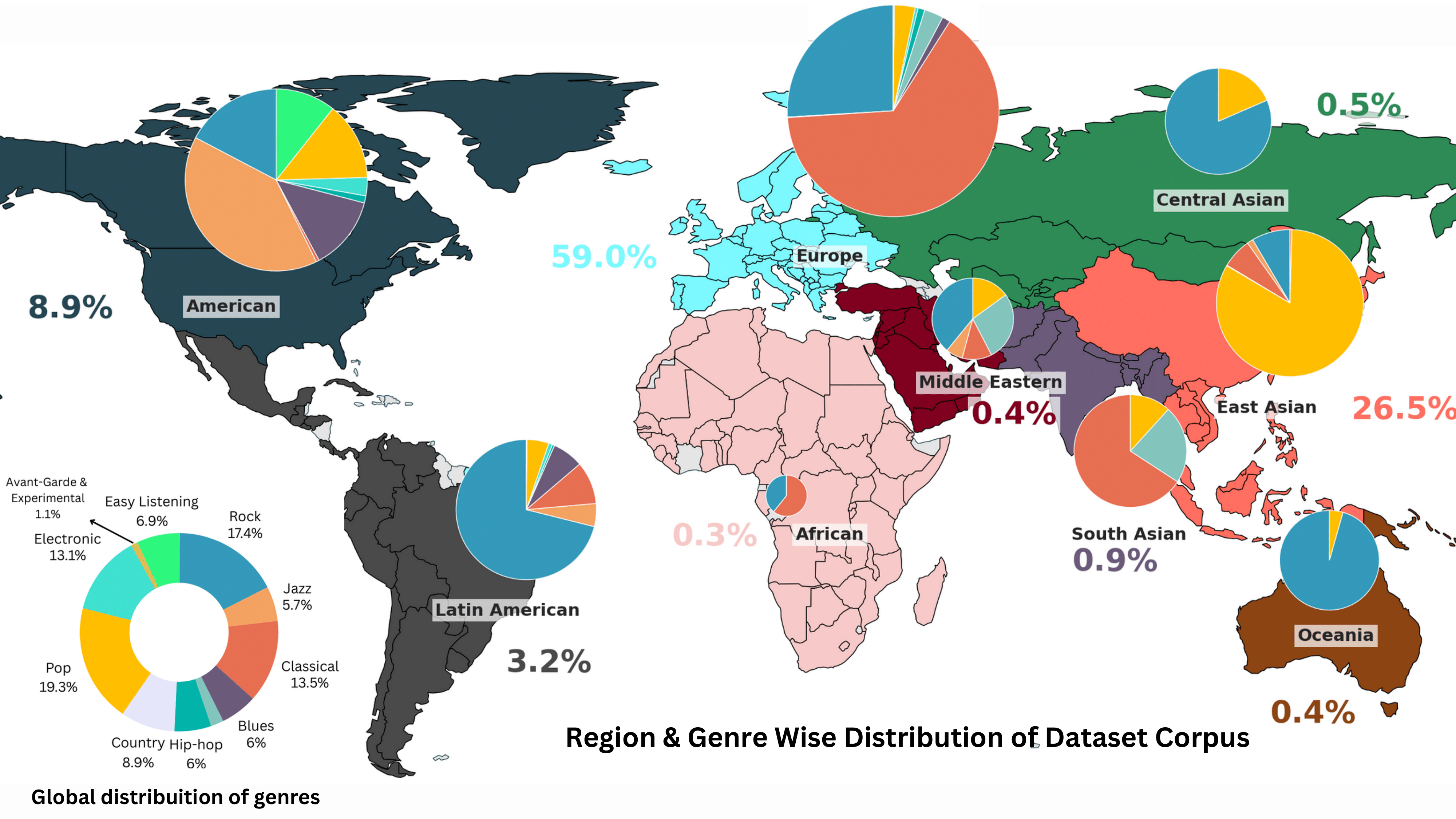}
\caption{%
The bottom left piechart shows the global distribution of genre. Each piechart in the map shows the distribution of genres in different regions with the size of each piechart being proportional to their contribution to the data corpus.
}
\label{fig:dataset_infographic}
\end{figure*}

We hypothesize that explicit fine-tuning with a small number of additional parameters (less than 1\% of the pre-trained model) as adapters~\cite{bapna2019simple}, will lead to a far better performance for the under-represented music style. We conduct a series of experiments comparing the performance of baseline models and our adapter-enhanced models on objective metrics. For human evaluations, each model is tested using our novel evaluation framework roughly based on Bloom’s Taxonomy \cite{armstrong2010}: \textbf{Recall, Analysis}, and \textbf{Creativity} evaluate a model’s audio generation. Recall reproduces trained entities, Analysis forms new combinations of them, and Creativity blends different entities across genres in novel, unseen ways. Evaluations are conducted in a play-arena style, ranking models based on their adherence to the text prompt in terms of \textit{rhythm}, \textit{instrument}, \textit{melody} and \textit{creativity}. Mustango shows improvement when finetuned on Hindustani Classical music by 8\% and MusicGen shows improvement by 4\% on Turkish Makam in ELO ratings from their respective baselines. 

Our results show that while PEFT techniques are effective in improving the overall quality of generated music for the under-represented genre over the baseline models, not all models are adaptable to all genres. This implies that the various design choices made in the architecture, and the training datasets and recipes for the base model are crucial determinants of the adaptability of a model to certain musical genres.

The contributions of this paper are threefold:

\begin{enumerate}
\setlength{\parskip}{0pt}
    \item We provide a detailed analysis of the current state of musical datasets, highlighting the under-representation of non-Western music.
    \item We present the first application of parameter-efficient training with adapters for cultural adaptation of under-represented genres in music generation models. 
    \item We introduce a novel arena-style evaluation framework based on Bloom's Taxonomy to assess the text to music generation capabilities of models using a play-arena style, ranking models on on their adherence to the text prompt on \textit{rhythm}, \textit{instrument}, \textit{melody} and \textit{creativity}.
    \item We demonstrate that while adapting base models to different genres is possible, it is a non-trivial challenge.
\end{enumerate}
The rest of the paper is organized as follows: In Section \ref{disparity_research}, we discuss global disparities in music representation, followed by Section \ref{experimental_setup}, which details our approach to adapting genres. Section \ref{result} presents our evaluation methodology and the results obtained from our analysis. We conclude our findings in Section \ref{conclusion}.

\section{The Disparity in Music Generation Research} \label{disparity_research}
AI music generation has evolved rapidly with techniques such as autoregressive~\cite{agostinelli2023musiclm,c:23,ziv2024masked}, diffusion-based ~\cite{c:24,huang2023noise2music,li2024jen} and GAN-based ~\cite{dong2018musegan,li2021inco} producing high-quality music. Some of the works include adapter-based settings which proved effective for music editing and inpainting~\cite{lin2024arrange, zhang2024instruct}. Moreover,~\citet{lan2024musicongen} used adapters for rhythm and chord conditioning. ~\citet{tan2020automated} showcased how visual emotions from images can be effectively translated into music using deep learning techniques.

Drawing inspiration~\citet{c:27}, which systematically analyzes the under-representation of languages spoken by the global majority, we conduct a survey of the datasets and research papers on music generation.

\subsection{Data Collection}
To get our initial pool of papers, we implemented an efficient, automated data collection method.

We employed a multi-stage, keyword-based selection method, leveraging the Scholarly package~\citet{cholewiak2021scholarly} to gather approximately 5000 papers. This included up to 1000 papers per query, using broad search terms such as “music,” “music generation,” “non-Western music,” “MIDI,” and “symbolic music.” We then refined our selection by focusing on papers presented at 10 major conferences including \textit{IJCAI, AAAI, ICML, EURASIP, EUSIPCO, ISMIR, NeurIPS, SMC, NIME} and \textit{ICASSP}, chosen based on their popularity and prestige in the area of computational processing of music, narrowing our pool to around 800 papers. Conferences such as ISMIR and NIME specialize in music information retrieval and musical expression, frequently showcasing work related to generative AI. Additionally, conferences like ICASSP, AAAI, and NeurIPS are known for their focus on cutting-edge AI technologies, such as GANs and transformers, which are crucial for music generation.

\subsubsection{Dataset Papers}
To identify papers proposing datasets, we read through the title and abstract of each paper. This led to a set of  \textbf{152} papers proposing new datasets with a total of \textbf{1 million+} hours of music. These datasets were manually annotated for the region and genres covered, total hours of music data, and whether the dataset is annotated with other details (such as, instruments, genre, and style). Papers that directly provided details of the distribution of data points across genres and regions, were analyzed with the already available statistics. Unfortunately, several datasets did not offer substantial details necessary for our study. If such a dataset had more than \textit{\textbf{10,000}} hours of audio data, we analyzed each sound file's metadata to collect genre and region information. However, when the genre and region were not explicitly mentioned in either the paper or the metadata, we did not make any assumptions; thus, \textbf{7.9\%} of the datasets totaling\textbf{ 5,772} hours were excluded from our analysis. 

\subsection{Findings}
Our findings are summarized in Figure \ref{fig:dataset_infographic}. The results reveal an almost complete omission of musical genres from non-Western countries, especially those from the Global South. Approximately 94\% of the total hours in available datasets are dedicated to music from the Western world, while only 5.7\% are devoted to \textit{South Asian, Middle Eastern, Oceanian, Central Asian, Latin American}, and \textit{African} music combined. This imbalance is likely to cause poor-quality music generation for genres from the Global South. For detailed analysis, please refer to Appendix \ref{appendix:research_landscape}.

\section{Genre Adaptation: Data, Models and Experiments} \label{experimental_setup}
For our genre-adaptation experiments, we selected two distinct non-Western genres — Hindustani Classical~\cite{jairazbhoy1971rāgs} and Turkish Makam~\cite{signell2008makam} — both significantly underrepresented in music generation research and datasets, and two open source models -- MusicGen~\cite{c:23} and Mustango~\cite{melechovsky-etal-2024-mustango}. We begin by describing the dataset creation, followed by prompt generation, the models, adapter architectures and finally, the training process.

\subsection{Dataset Creation}

Our study necessitated diverse corpus of non-Western music with detailed metadata. The Dunya~\cite{porter-2013} which is part of the CompMusic project~\cite{serra2014creating}, emerged as the ideal choice, offering an extensive collection of over 1,300 hours of music across multiple non-Western genres. This corpus includes Carnatic, Hindustani, Turkish Makam, Beijing opera, and Arab Andalusian music, providing a broad spectrum of cultural music. We focused specifically on Hindustani Classical and Turkish Makam genres as both genres possess complex culturally specific melodic and rhythmic structures different from Western music \& we had easier access to listeners familiar with Indian and Turkish music. For Hindustani Classical, we chose the MTG Saraga~\cite{srinivasamurthy2021saraga} annotated dataset which is built on CompMusic offering 50 hours of audio. For Turkish Makam, we use the Dunya dataset API for accessing the metadata and audio samples leading to 405 hours of audio. 

To ensure consistency and manage computational resources effectively, we implemented several pre-processing steps. We standardized the audio sample length by truncating longer recordings to 30 seconds. We utilized the accompanying metadata from the Dunya corpus without modification. These descriptions, rich in genre-specific details, served as valuable inputs for creating prompt templates. Finally, to accommodate the differing requirements of our chosen models, we performed audio resampling. Specifically, for MusicGen we resampled the audio to a 32 kHz sampling rate and for Mustango 16 kHz sampling rate.

The metadata from the dataset provides genre-specific information for each audio clip, including three key details critical to our study: melodic line, rhythmic pattern, and instrumentation. For the melodic line, we extracted the \textit{raga} (a melodic framework in Hindustani Classical music) and \textit{Makam} (a system of melodic modes in Turkish music). For rhythmic patterns, we identified \textit{laya} (tempo) in Indian music and \textit{usul} (a sequence of rhythmic strokes) in Turkish music. Additionally, we extracted the meta-data for the instruments (including voice) played in each audio sample. Details of the dataset can be found in Appendix \ref{appendix:dataset_details}.


\begin{figure*}[!t]
    \centering
    \includegraphics[width=1\linewidth]{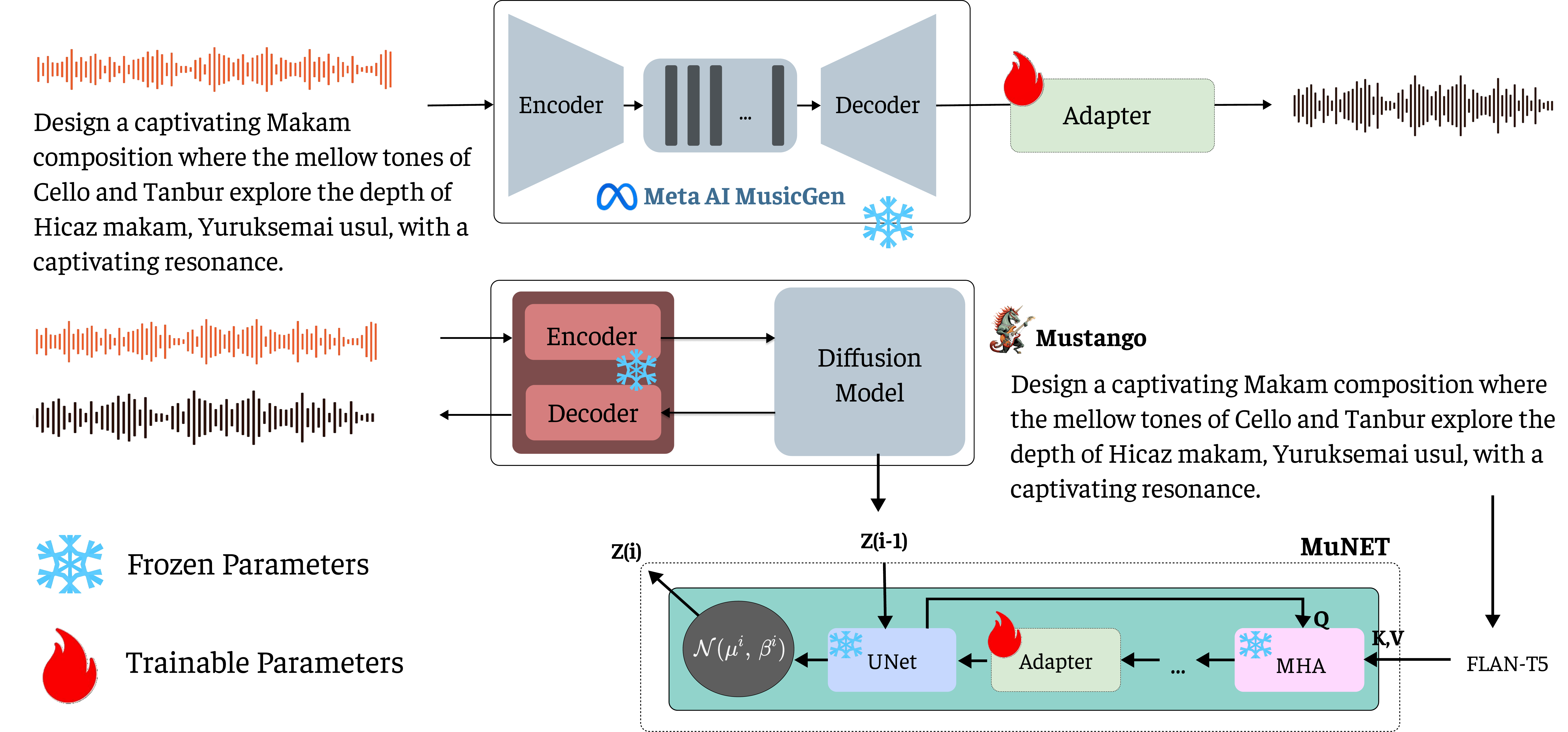}
    \caption{Mustango \& MusicGen settings for low-resource fine-tuning.}
    \label{fig:architecture}
\end{figure*}

After pre-processing, we collected a total of 23.24 hours of audio for Hindustani Classical music and 121.16 hours for Turkish Makam music. The dataset was then divided using an 80-20\% split for training and testing, allowing us to evaluate the final model performance effectively. We ensured that audio clips for training and testing come from different songs to prevent distribution overlap in train and test. This split resulted in 18.91 hours of Hindustani Classical music and 97.23 hours of Turkish Makam music for training, and the remaining portions reserved for testing.

\subsection{Prompt Generation}

To create effective prompts for model training, we created three distinct templates that describe each musical piece based on sample metadata from the selected genres. 

\begin{table}[!t]
\centering
\small
\setlength{\tabcolsep}{6pt} 
\begin{tabularx}{\columnwidth}{lX} 
\toprule
\textbf{Query Type} & \textbf{Example} \\
\midrule
\textbf{Recall} & Imagine a traditional $\bigstar$ \textbf{Makam} performance that brings together $\triangleright$ Clarinet, $\triangleright$ Darbuka, $\triangleright$ Kanun, $\triangleright$ Oud, $\triangleright$ Voice, $\sharp$ Aksak makam, and $\flat$ Hicaz usul, flowing effortlessly. \\ \midrule
\textbf{Analysis} & Imagine a traditional $\bigstar$ \textbf{Makam} performance that brings together $\triangleright$ Tanbur, $\triangleright$ Oud, $\triangleright$ Cello, with the flowing essence of $\sharp$ Aksak makam and $\flat$ Fahte usul, flowing effortlessly. \\ \midrule
\textbf{Creativity} & Imagine a modern $\bigstar$ \textbf{Western Electronic Dance Music (EDM)} performance infused with the soulful sound of $\triangleright$ Tanbur, rich vocals blending with $\sharp$ Acem makam and $\flat$ Fahte usul. \\
\bottomrule
\end{tabularx}
\caption{Recall, Analysis \& Creativity Queries: Recall uses known combinations, while Analysis introduces novel combinations to test analytical capability and Creativity introduces cross-genre combinations. Refer to Section \ref{result}. \textbf{Genre}:$\bigstar$, melodic line:$\sharp$, rhythmic pattern:$\flat$, and instrumentation:$\triangleright$.}
\label{tab:queries}
\end{table}

For each audio sample, we randomly selected one of the three templates and populated it with relevant metadata attributes as shown in Table \ref{tab:queries}.

\noindent This process ensures that each prompt captures the unique musical elements of the sample. By maintaining this metadata-specific structure across prompts, we help the model learn to identify and respond to key attributes within each genre, enabling it to generate more accurate and culturally informed outputs during training.

\subsection{Models}
We utilize two state-of-the-art models, MusicGen~\cite{c:23} and Mustango~\cite{melechovsky-etal-2024-mustango}, to explore cross-genre adaptation. MusicGen is a transformer-based model, while Mustango integrates both diffusion and transformer architectures. We introduce adapters~\cite{pfeiffer-etal-2020-adapterhub} that enable low-resource fine-tuning. We also considered Moûsai \cite{c:24} and MusicLM \cite{agostinelli2023musiclm}, but Moûsai and MusicLM lack open-source training codes.

\subsubsection{MusicGen}
In MusicGen, we enhance the model with an additional 2 million parameters by integrating \textbf{Bottleneck Residual Adapter} after the transformer decoder within the MusicGen architecture after thorough experimentation with other placements. The total parameter count of MusicGen is 2 billion, making the adapter only 0.1\% of the total size. The adapter, as shown in Figure \ref{fig:architecture}, consists of a linear layer that compresses the embedding to a very small dimension, followed by a non-linear activation and projection back to the original size.

MusicGen leverages the Encodec~\cite{defossez2022highfi} framework, which compresses audio into latent representations. These latent representations are processed through a transformer model, which generates new music based on input prompts. By placing adapters at the end of the decoder, we achieve a lightweight adaptation mechanism that enhances the model’s ability to generate music in specific styles or regions, such as Hindustani Classical music and Makam, without modifying the fundamental Encodec structure.
\vspace{-1pt}
\subsubsection{Mustango}
In Mustango, we enhance the model with an additional 2 million parameters, which represents only 0.1\% of the model's total parameter count, by integrating a \textbf{Bottleneck Residual Adapter}. 

While Mustango supports chord and beat embeddings, we opted not to use them here due to the distinct focus of Hindustani Classical and Turkish Makam on melodic lines rather than harmonic progressions. Unlike Western classical music, these genres feature complex, rhythms with accents often within a single beat, making fixed beat and chord embeddings difficult to apply.

The adaptation process in Mustango begins with the FLAN-T5~\cite{flan-t5} model, which converts the input text into embeddings. These embeddings are then incorporated into the UNet architecture~\cite{10.1007/978-3-319-24574-4_28} through a cross-attention mechanism, aligning the text and audio components. To refine this process, a Bottleneck Residual Adapter with convolution layers is incorporated into the up-sampling, middle, and down-sampling blocks of the UNet, positioned immediately after the cross-attention block at the end of each stage (Figure \ref{fig:architecture}). The adapters reduce channel dimensions by a factor of 8, using a kernel size of 1 and GeLU activation after the down-projection layers to introduce non-linearity. Various adapter configurations and placements were explored to preserve the musical structure while adapting stylistic elements, with this setup yielding the best output quality. This design facilitates cultural adaptation while preserving computational efficiency. 

\subsection{Training settings}
For MusicGen fine-tuning, we used two RTX A6000 GPUs over a period of around 10 hours. The adapter block was fine-tuned, using the AdamW~\cite{Loshchilov2017DecoupledWD} optimizer with a learning rate of 5e-5 and a weight decay of 0.05 using MSE based Reconstruction Loss. The training spanned 20 epochs, with a patience threshold of 5 epochs for early stopping based on validation loss. We utilized a batch size of 4 and applied gradient clipping with a maximum norm of 1.0. The training data was split into 90\% for training and 10\% for validation.

For Mustango model fine-tuning, we used one RTX A6000 GPU over a period of 12 hours.  The adapter block was fine-tuned, using the AdamW optimizer with a learning rate of 4.5e-5 and a weight decay of 0.01 using MSE based Reconstruction Loss. The training spanned 25 epochs for both genres, with a patience threshold of 5 epochs for early stopping based on validation loss. The training data was split into 80\% for training and 20\% for validation. 
\section{Results} \label{result}
\begin{table}[!t]
\centering
\small 
\setlength{\tabcolsep}{10pt} 
\begin{tabular}{lcccc}
\toprule
\multicolumn{5}{c}{\textbf{Objective Metrics}} \\ \midrule
\multicolumn{5}{c}{Hindustani Classical Music} \\ \midrule
\textbf{Model} & \textbf{FAD ↓} & \textbf{FD ↓} & \textbf{KLD ↓} & \textbf{PSNR ↑} \\ \midrule
\textbf{MGB} & 40.05 & 75.76 & 6.53 & 16.23 \\
\textbf{MGF} & 40.04 & 72.65 & 6.12 & 16.18 \\
\textbf{MTB} & 6.36 & 45.31 & 2.73 & 16.78 \\
\textbf{MTF} & \textbf{5.18} & \textbf{22.03} & \textbf{1.26} & \textbf{17.70}\\ \midrule

\multicolumn{5}{c}{Turkish Makam} \\ \midrule
\textbf{Model} & \textbf{FAD ↓} & \textbf{FD ↓} & \textbf{KLD ↓} & \textbf{PSNR ↑} \\ \midrule
\textbf{MGB} & 39.65 & 57.29 & 7.35 & 14.60 \\
\textbf{MGF} & 39.68 & 56.71 & 7.21 & 14.46 \\
\textbf{MTB} & 8.65 & 75.21 & 6.01 & \textbf{16.60} \\
\textbf{MTF} & \textbf{2.57} & \textbf{20.56} & \textbf{4.81} & 16.17\\ 
\bottomrule
\end{tabular}
\caption{Objective Evaluation Metrics for Hindustani Classical Music and Turkish Makam.}
\label{tab:objective_model_metrics}
\end{table}

\begin{figure*}[!t]
    \centering
    \includegraphics[width=\textwidth]{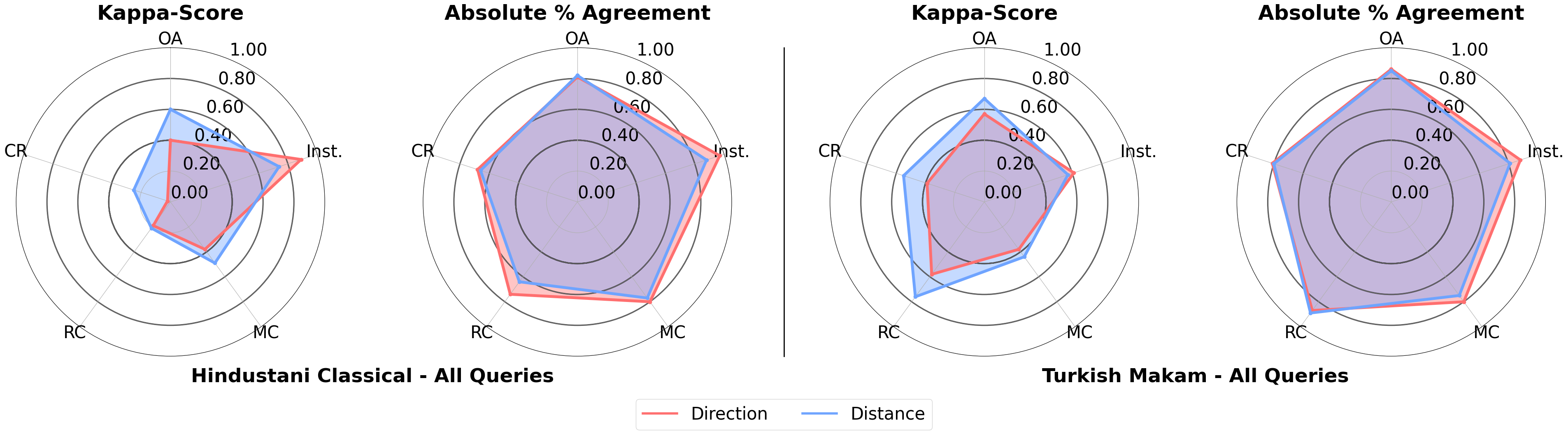} 
    \caption{Inter Annotator Agreement metrics for Hindustani Classical \& Turkish Makam.}
    \label{iaa_evaluation}
\end{figure*}
We evaluated four models, Mustango Baseline (MTB), Mustangto Fine-tuned (MTF), MusicGen Baseline (MGB), and MusicGen Finetuned (MGF), on two genres using both objective metrics and human evaluation, providing both objective and subjective insights into model performance. 
\subsection{Automatic Metrics}
We sample 400 audio samples from the test set to form our test prompt corpus. For capturing the distance between generated audio and the test corpus we compute Fréchet Audio Distance(FAD), Fréchet Distance(FD) and Kullback-Leibler(KL) with Sigmoid activation. We utilize the AudioLDM~\cite{liu2023audioldm} toolkit for implementation of FAD, FD, and KL, with distributions computed using PANN-CNN14~\cite{Kong_Cao_Iqbal_Wang_Wang_Plumbley_2020} as the backbone model for extracting features for each audio sample. 



 Table~\ref{tab:objective_model_metrics} presents the performance metrics for models trained on Hindustani Classical music. Both finetuned versions of MusicGen and Mustango demonstrate superior performance compared to their baseline counterparts across all evaluated metrics. Notably, Mustango exhibits significant improvement after finetuning, whereas MusicGen shows only marginal gains. This disparity suggests that Mustango possesses a greater capacity for dataset-specific adaptation.

Adapter finetuning for Mustango model better incorporates domain-specific nuances of Hindustani Classical music, resulting in generated outputs that more closely align with the target style.


Table~\ref{tab:objective_model_metrics} additionally presents results for Turkish Makam generation. The performance trends mirror those observed in Hindustani Classical music. Mustango demonstrates strong improvement with one notable exception: the PSNR metric. For PSNR, the baseline Mustango model performs better.

\subsection{Human Evaluation}
To complement our objective metrics, we designed a rigorous human evaluation process, recognizing the crucial role of human perception in assessing music quality and authenticity. We begin by generating prompts for drawing audio inferences from the models based on Bloom's taxonomy criteria. Then we present the outputs to human judges to compare them in an arena setup~\cite{chiang2024chatbot, tts-arena}.  

We divided our process into two phases. In first phase, two annotators independently judged a portion of the same set of data points. This allowed us to compute inter-annotator agreement, a crucial measure of evaluation reliability. Disagreements were systematically discussed and resolved, refining our evaluation criteria. In phase two, annotators transitioned to single annotations per data point continuing evaluation of the rest audios. Finally, we compute ELO ratings of the models based on second phase annotations. 

\subsubsection{Material}

We introduce novel evaluation criteria based on Bloom's Taxonomy to assess a model's understanding of musical elements in text and their alignment with the generated audio using arena-style evaluation.

We evaluate the models under three conditions: \textbf{recall}, \textbf{analysis}, and \textbf{creativity}, by manually generating 10 prompts in each category~(Table \ref{tab:queries}).
\begin{itemize}
\setlength{\parskip}{0pt}
    \item \textbf{Recall}: Tests the model’s ability to reproduce combinations of melody, instrument, and rhythm from the fine-tuning data, testing effective memorization and recall.


    \item \textbf{Analysis}, We create novel combinations by substituting melodies, rhythms, or instruments, testing the model's adaptability beyond the training data.

    
    \item \textbf{Creativity}, We combine genres, blending melodies, rhythms, and instruments across styles to test the model's integration of underrepresented and over-represented genres.
    

    
\end{itemize}

\noindent For each case, we generate model responses from all models, creating 120 total music samples. Since Mustango generates 10-second inferences at 16kHz, we process MusicGen outputs by clipping them to 10 seconds and downsampling to 16kHz to ensure uniform evaluation conditions.

\subsubsection{Method}

\begin{figure*}[!t]
    \centering
    \includegraphics[width=0.98\textwidth]{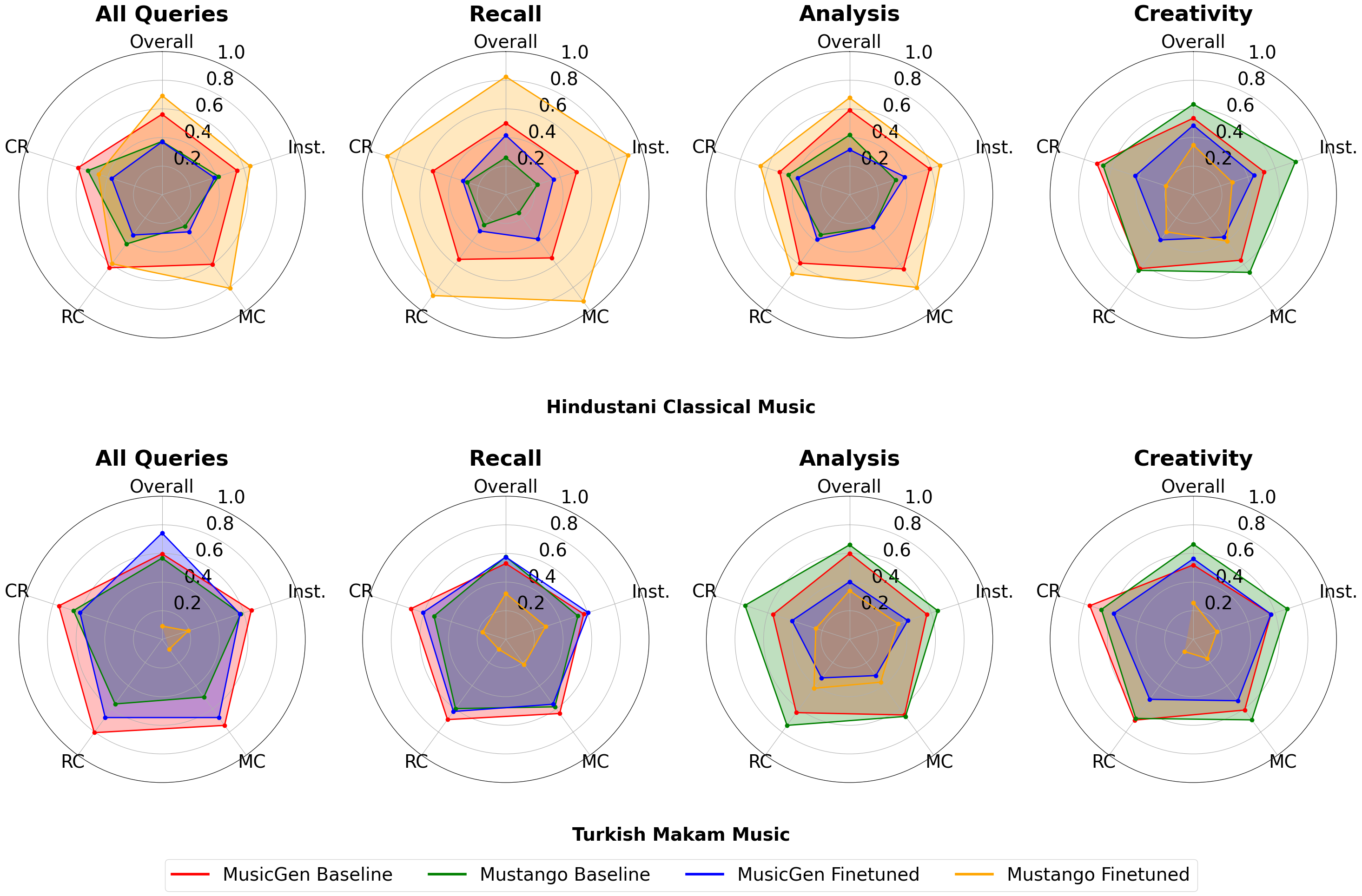} 
    \caption{Scaled ELO ratings for each model in Hindustani Classical and Turkish Makam music. Categories include OA: Overall Assessment, Inst.: Instrument Accuracy, MC: Melody Capture, RC: Rhythm Capture, and CR: Creativity. Ratings are provided for all query types and individual query categories.}
    \label{fig:results_evaluation}
\end{figure*}

We decided to go for a comparative evaluation of pieces instead of absolute judgments of pieces in isolation to control the subjectivity so that, the shorter, lower-sampling-rate music clips, are more effectively evaluated through comparison. For each comparison, the user receives a reference prompt and two anonymous audio samples (with the comparisons ordered randomly), followed by five comparative evaluation questions comparing the two audio generations on each criterion: Overall Aesthetic\textit{(OA)}, Instrument Accuracy\textit{(Inst.)}, Rhythm Capture\textit{(RC)}, Melody Capture\textit{(MC)}, and Creativity\textit{(CR)} since we provide these entities in the prompt and we are trying to assess the alignment of the text to the audio generated. For each criteria, we provide the annotator with 7 options: A $\gg$ B, A $>$ B, A = B, A $<$ B, A $\ll$ B, None, and Not Applicable (NA). Please refer to Appendix~\ref{appendix:annotation_det} for questions.

\textbf{In first phase,} we conduct four types of comparisons: \textit{baseline} vs. \textit{baseline},\textit{ baseline }vs.\textit{ fine-tuned} (for both models), and \textit{fine-tuned} vs. \textit{fine-tuned}. We request two avid listeners of each genre, who are aware of the nuances but not themselves professional musicians(demography details in Appendix \ref{appendix:annotator_demographics}), to annotate these samples. The annotation process begins by evaluating 36 comparisons for each genre—9 generations from each model per genre—compared across all models based on the five evaluation criteria. After the completion of first phase we compute the \textbf{Inter-Annotator Agreement(IAA)}, using \textbf{distance} and \textbf{direction-based kappa} scores. The distance-based Kappa quantifies the absolute differences in annotations by both the annotators whereas direction-based Kappa assesses consistency in preference order rather than the extent of preference. Detailed kappa-score calculation methods are provided in the Appendix \ref{appendix:kappa_score_compute}. 





Figure \ref{iaa_evaluation} presents the kappa score and average agreement for each criterion.
In evaluating Hindustani Classical and Turkish Makam music, these metrics reveal distinct patterns across assessment criteria. As we can see from the figure, \textit{OA} scores range from 0.40 to 0.67, while \textit{Inst.} shows high agreement, with scores up to 0.89 due to its objectivity (Figure \ref{iaa_evaluation}). \textit{MC} achieves moderate agreement, and \textit{RC} scores are generally lower, particularly for Hindustani Classical (0.19 and 0.21), likely due to the complexity of rhythmic patterns within the short 10-second evaluation span. \textit{CR} consistently records the lowest scores, reflecting the subjective nature of this criterion(For more genre-wise and query-type-wise details refer to Appendix\ref{appendix:iaa}). After phase-I, we ask the annotators to discuss and re-annotate music samples for Inst. and RC criteria where disagreement is higher.

\textbf{In the second phase,} for Hindustani Classical, we removed the MGB vs MGF comparison since the trend made it clear that MGB is better with agreement from both the annotators. For Turkish Makam, we removed MTF vs MTB since MTB proved to be better(see Figure \ref{fig:results_evaluation}). After filtering we are left with 3 sets of comparisons, with 7 queries for each model, across 3 query types leading to 63 more comparisons for each genre. The annotations from both rounds are combined to compute each model’s ELO rating.

\subsubsection{ELO Ratings}
After comparing the model outputs, we compute ELO ratings (usually used to calculate the relative skill levels of players in a two-player game) for each model across all query types for each evaluation criterion.
For each criterion, we consider a single annotation as a single match between the models. If the annotator marks it as NA, then we omit it from the calculation, if A$=$B or None is marked we consider it as a draw and A$\gg$B, A$>$ is considered a win for A and vice-versa for the remaining cases. The details of computing ELO ratings are given in Appendix \ref{appendix:elo}.

The normalized ELO ratings are shown in Figure~\ref{fig:results_evaluation}. \textbf{For Hindustani Classical music}, over all queries, MTF outperforms all models. Interestingly, MGB is better than MTF, but MGF is judged least favourably, implying that while fine-tuning significantly improves Mustango, MusicGen's performance regresses considerably. These trends hold for all aspects (melody, rhythm, instrument) except creativity. The trends are most prominent for Recall queries, but also hold for Analysis queries, but completely reverses for creativity queries, where MTF performs the worst while MTB performs the best. Qualitative analysis of the generated pieces confirm this finding and shows that there was a strong effect of adaptation on Mustango which led to knowledge attrition and resultant poor performance on creative queries, which required the model to utilize previous knowledge.

For Turkish Makam music, MTF regresses significantly from MTB, for types of queries as well as on all aspects. While MGF performs slightly better than MGB on all queries for overall rating, the trends are not consistent across different aspects, or different types of queries. In fact, for Analysis queries, even MusicGen's performance significantly regresses for all aspects on finetuning. Thus, we can conclude that the PEFT technique explored here did not help boost the performance of the models for Turkish Makam music. 


\section{Conclusion} \label{conclusion}
In this paper, for the first time, we systematically explored and established the skewed distribution of musical genres from around the world in datasets used for training Music-Language Models. Non-Western musical traditions are severely underrerpresented which naturally leads to disparate performance across genres in these models. We also demonstrate that PEFT-based techniques vary in effectiveness across different genres and models, further aggravating the challenges of overcoming the data scarcity problem.

As generative models continue to gain traction in the field of music generation and are expected to be used even more widely in the coming years, the misrepresentation and under-representation of the musical genres of the ``global majority" poses a significant threat to the inclusion of musical cultures from around the world. The skewed distribution in datasets, reflected in model outputs, can lead to several issues, including cultural homogenization, reinforcement of Western culture dominance \cite{crawford2016}, misrepresentation of musical styles, and most importantly, gradual decline leading to the disappearance of many musical genres~\cite{tan-2021,decolon_pop,intercontinental2023ai}. Therefore, it is critically important to prioritize the creation of inclusive music datasets and models, with an emphasis on under-represented musical genres.


\newpage
\section*{Limitations}
Our work relies on adapter-based techniques for cross-cultural adaptation but there is a need to explore additional architectural configurations to further optimize low-resource fine-tuning such as LoRA~\cite{DBLP:journals/corr/abs-2106-09685} or Compacter~\cite{Davison2021CompacterEL} approaches. 

Additionally, our approach only focused on a few genres, and future work should aim to incorporate a broader range of musical styles. Our investigation involves only Hindustani classical and Turkish Makam traditions, leaving other genres from the Dunya dataset unexplored. This narrow focus stems not from a lack of curiosity, but from our limited cultural expertise - a constraint we acknowledge upfront.

We also trained separate models for Hindustani Classical and Turkish Makam music; combining these into a single model could offer greater generalization across genres. 

Another limitation lies in the evaluation process. Human evaluations were conducted on a limited number of samples with a duration of 10 seconds, and more genre-specific assessments are necessary. We also believe that computing objective metrics for underrepresented genres may obscure the full picture because the backbone models used to compute these metrics may not have been trained on various underrepresented genres, resulting in an erroneous portrayal of genres.

\bibliography{custom}

\appendix
\appendix 


\section{Research Landscape} \label{appendix:research_landscape}
\begin{table}[htbp]
\centering
\resizebox{\columnwidth}{!}{
\begin{tabular}{lrr}
\toprule
\textbf{Region} & \textbf{Papers (count)} & \textbf{Duration (hrs.)} \\ \midrule 
\textbf{European}    & $66$ & $\mathbf{6127.92}$ \\ \midrule
East Asian           & $\mathbf{71}$ & $2746.73$ \\ 
South Asian          & $1$ & $88.78$ \\
Central Asian        & $0$ & $57.01$ \\ \midrule
American             & $72$ & $921.84$ \\ 
Latin American       & $5$ & $323.25$ \\ \midrule
Oceania              & $3$ & $41.99$ \\ \midrule
African              & $0$ & $27.50$ \\ \midrule
Middle Eastern       & $5$ & $37.86$ \\ \bottomrule
\end{tabular}}
\caption{Distribution of Papers and Duration in hours by Region.}
\label{tab:region_table}
\end{table}

\subsection{Genre Distribution Analysis}

Genre-wise analysis we can observe in Table \ref{tab:genre_table} and that \textit{Pop} music forms 200K+ hours of the corpus while \textit{Folk} music constitutes only 20K hours of the corpus which is a 10 times between the two as shown in Table \ref{tab:genre_table}. \textit{Pop, Rock, Classical} and \textit{Electronic} music genres each constitute more than 10\% of the total corpus and more than 100K hours in the total corpus. As shown in Figure \ref{fig:dataset_infographic}, \textit{Pop} music has the highest (19.3\%) representation followed by \textit{Rock}~(17.4\%) and \textit{Classical}~(13.5\%) genres. \textit{Country, hip-hop}\textit{, blues }and \textit{jazz} have a moderate~(more than 5\%) representation in the corpus. \textit{Folk} and \textit{experimental} music contribute to only 2.1\% of the corpus. The other genres receive minimal attention(\textbf{$\leq 1\%$}) which includes music for Children, \textit{Indie-music}, and region-specific genres.
 
\subsection{Regional Distribution Analysis}
In region-wise analysis, from analyzing the research space we find that more than 6k hours of music in the research belongs to \textit{European} music and only 28 hours of music belong to \textit{African} music as shown in Table \ref{tab:region_table}. \textit{European}, \textit{East Asian} and\textit{ American} music are well explored forming 84.5\% of the corpus. On the other hand, \textit{South Asian, Middle Eastern, Central Asian} and \textit{African} music each contribute less than 1\% to the whole corpus as depicted in Figure \ref{fig:dataset_infographic}.

\begin{table}[htbp]
\centering
\resizebox{0.9\columnwidth}{!}{
\begin{tabular}{lrr}
\toprule
\textbf{Genre} & \textbf{Papers (count)} & \textbf{Duration (hrs.)} \\ \midrule
\textbf{Pop} & \textbf{24} & \textbf{206.89} \\
Rock & 7 & 186.67 \\
Electronic & 36 & 140.25 \\ \midrule
Classical & 91 & 144.64 \\
Country & - & 95.77 \\
Hip-hop & 3 & 64.35 \\ \midrule
Jazz & 15 & 60.62 \\
Blues & - & 64.01 \\
Easy Listening & 2 & 74.39 \\ \midrule
Folk & 3 & 22.802 \\ 
Experimental & 26 & 11.310 \\
Others & 15 & 0.94 \\
\bottomrule
\end{tabular}}
\caption{Distribution of Hours and Papers by Genre. Duration (Dur.) is represented as 10\textsuperscript{3} hours.}
\label{tab:genre_table}
\end{table}

\section{Annotation Details} \label{appendix:annotation_det}
We asked annotators to choose between two audio samples, based on their preference, to select which better represents the prompted culture in both the inter-annotator agreement scenario and human evaluation. For both inter-annotator agreement and human evaluation, we relied on the same set of questions outlined below.
\begin{itemize}
\setlength{\parskip}{0pt}
    \item Overall, which piece do you like more?
    \item Which piece captures the instrument (if mentioned the prompt) better?
    \item Which piece captures the melodic line/scale (if mentioned the prompt) better?
    \item Which piece captures the rhythm/tempo (if mentioned the prompt) better?
    \item Which piece is more creative (ignore audio quality while answering this question)?
\end{itemize}

\section{Evaluation of Inter Annotator Agreement Results} \label{appendix:iaa}
\begin{table*}[!t]
\begin{minipage}{\textwidth}
\centering
\small 
\setlength{\tabcolsep}{20pt} 
\resizebox{0.75\textwidth}{!}{
\begin{tabular}{lccccc}
\toprule
\multicolumn{6}{c}{\textbf{Inter Annotator Agreement (Kappa Score, ↑)}} \\ \midrule

\multicolumn{6}{c}{Hindustani Classical - All Queries} \\ \midrule
\textbf{Metric Type} & \textbf{OA} & \textbf{Inst.} & \textbf{MC} & \textbf{RC} & \textbf{CR} \\ \midrule
Direction & 0.40 & 0.89 & 0.38 & 0.19 & 0.02 \\
Distance  & 0.60 & 0.74 & 0.49 & 0.21 & 0.25 \\ \midrule

\multicolumn{6}{c}{Turkish Makam - All Queries} \\ \midrule
\textbf{Metric Type} & \textbf{OA} & \textbf{Inst.} & \textbf{MC} & \textbf{RC} & \textbf{CR} \\ \midrule
Direction & 0.57 & 0.61 & 0.38 & 0.58 & 0.39 \\
Distance  & 0.67 & 0.57 & 0.44 & 0.76 & 0.55 \\
\bottomrule
\end{tabular}}
\caption{Inter Annotator Agreement for Hindustani Classical and Turkish Makam genres. The IAA is calculated using both Direction and Distance-based metrics.}
\label{tab:inter_annotator_agreement}

\centering
\small 
\setlength{\tabcolsep}{20pt}
\resizebox{0.75\textwidth}{!}{
\begin{tabular}{lccccc}
\toprule
 \multicolumn{6}{c}{\textbf{Inter Annotator Agreement (Kappa Score, ↑)}} \\ \midrule

\multicolumn{6}{c}{\textbf{Hindustani Classical Music}} \\ \midrule

\multicolumn{6}{c}{Recall Queries} \\ \midrule
\textbf{Metric Type} & \textbf{OA} & \textbf{Inst.} & \textbf{MC} & \textbf{RC} & \textbf{CR} \\ \midrule
\textbf{Direction} & 0.38 & 1 & 0.07 & 0.72 & -0.04 \\
\textbf{Distance}  & 0.48 & 0.63 & 0.33 & 0.39 & 0.32 \\ \midrule

\multicolumn{6}{c}{Analysis Queries} \\ \midrule
\textbf{Metric Type} & \textbf{OA} & \textbf{Inst.} & \textbf{MC} & \textbf{RC} & \textbf{CR} \\ \midrule
\textbf{Direction} & 0.43 & 1 & 0.48 & -0.56 & 0.15 \\
\textbf{Distance}  & 0.66 & 1 & 0.63 & -0.11 & 0.33 \\ \midrule

\multicolumn{6}{c}{Creativity Queries} \\ \midrule
\textbf{Metric Type} & \textbf{OA} & \textbf{Inst.} & \textbf{MC} & \textbf{RC} & \textbf{CR} \\ \midrule
\textbf{Direction} & 0.38 & 0.65 & 1 & 0.48 & -0.14 \\
\textbf{Distance}  & 0.63 & 0.59 & 0.63 & 0.38 & 0.06 \\ \midrule

\multicolumn{6}{c}{\textbf{Turkish Makam}} \\ \midrule

\multicolumn{6}{c}{Recall Queries} \\ \midrule
\textbf{Metric Type} & \textbf{OA} & \textbf{Inst.} & \textbf{MC} & \textbf{RC} & \textbf{CR} \\ \midrule
\textbf{Direction} & 0.74 & 1 & 0.38 & 0.65 & 0.74 \\
\textbf{Distance}  & 0.75 & 0.79 & 0.63 & 0.84 & 0.75 \\ \midrule

\multicolumn{6}{c}{Analysis Queries} \\ \midrule
\textbf{Metric Type} & \textbf{OA} & \textbf{Inst.} & \textbf{MC} & \textbf{RC} & \textbf{CR} \\ \midrule
\textbf{Direction} & 0.48 & 0.38 & -0.04 & 0.72 & 0.48 \\
\textbf{Distance}  & 0.63 & 0.33 & 0.18 & 0.80 & 0.45 \\ \midrule

\multicolumn{6}{c}{Creativity Queries} \\ \midrule
\textbf{Metric Type} & \textbf{OA} & \textbf{Inst.} & \textbf{MC} & \textbf{RC} & \textbf{CR} \\ \midrule
\textbf{Direction} & 0.48 & 0.55 & 1 & 0.38 & -0.04 \\
\textbf{Distance}  & 0.63 & 0.68 & 0.51 & 0.63 & 0.45 \\
\bottomrule
\end{tabular}}
\end{minipage}
\caption{Inter Annotator Agreement (IAA) Metrics across Recall, Analysis, and Creativity Queries for Hindustani Classical Music and Turkish Makam. Higher Kappa Scores (↑) indicate better agreement.}
\label{tab:inter_annotator_agreement_query}
\end{table*}
The inter-annotator agreement (IAA) results, measured using Cohen's Kappa, reveal interesting patterns across genres, metrics, and query types. 

In Table~\ref{tab:inter_annotator_agreement} Turkish Makam consistently showed higher agreement (0.57-0.67) than Hindustani Classical (0.40-0.60), suggesting potentially clearer structural elements. This trend is particularly pronounced in Rhythm (RC) annotations, where Turkish Makam exhibits substantially higher agreement (0.58-0.76) compared to Hindustani Classical (0.19-0.21).


Instrument identification (Inst.) showed high agreement across both genres (0.57-0.89), with Hindustani Classical scoring particularly well (0.89 for direction). Creativity (CR) exhibited the lowest overall agreement (0.02-0.55), reflecting the inherent subjectivity in assessing creativity.


Examining query types in Table~\ref{tab:inter_annotator_agreement_query} reveals that Recall queries generally yielded higher agreement, particularly in Turkish Makam (0.74-0.75). This indicates strong consistency in factual recall tasks. Analysis queries showed mixed results, with some categories in Hindustani Classical even showing negative agreement, pointing to potential confusion or divergent interpretations in analytical tasks. Interestingly, Creativity queries showed perfect agreement (1.0) in Melody for both genres, suggesting a strong consensus in perceiving creative aspects of melody.

\section{Evaluation of Human Evaluation Results} \label{appendix:human_eval_result}
\begin{table*}

\centering
\small 
\setlength{\tabcolsep}{20pt} 
\resizebox{0.76\textwidth}{!}{
\begin{tabular}{lccccc}
\toprule
\multicolumn{6}{c}{\textbf{Human Evaluation (ELO Ratings, ↑)}} \\ \midrule

\multicolumn{6}{c}{Hindustani Classical Music - All Queries} \\ \midrule
\textbf{Model} & \textbf{OA} & \textbf{Inst.} & \textbf{MC} & \textbf{RC} & \textbf{CR} \\ \midrule
\textbf{MusicGen Baseline} & 1525 & 1520 & 1540 & 1552 & 1546 \\
\textbf{Mustango Baseline} & 1449 & 1466 & 1409 & 1470 & 1518 \\
\textbf{MusicGen Finetuned} & 1448 & 1454 & 1428 & 1439 & 1448 \\
\textbf{Mustango Finetuned} & 1577 & 1559 & 1623 & 1538 & 1487 \\ \midrule

\multicolumn{6}{c}{Turkish Makam - All Queries} \\ \midrule
\textbf{Model} & \textbf{OA} & \textbf{Inst.} & \textbf{MC} & \textbf{RC} & \textbf{CR} \\ \midrule
\textbf{MusicGen Baseline} & 1539 & 1562 & 1597 & 1622 & 1603 \\
\textbf{Mustango Baseline} & 1527 & 1531 & 1499 & 1523 & 1560 \\
\textbf{MusicGen Finetuned} & 1597 & 1529 & 1570 & 1570 & 1541 \\
\textbf{Mustango Finetuned} & 1337 & 1377 & 1334 & 1286 & 1297 \\
\bottomrule
\end{tabular}}
\caption{Overall Evaluation Metrics for Hindustani Classical Music and Turkish Makam. ELO ratings (human evaluation) have higher values as better (↑).}
\label{tab:overall_model_metrics}

\centering
\small 
\setlength{\tabcolsep}{20pt} 
\renewcommand{\arraystretch}{1.1} 
\resizebox{0.76\textwidth}{!}{
\begin{tabular}{lccccc}
\toprule
\multicolumn{6}{c}{\textbf{Human Evaluation (ELO Ratings, ↑)}} \\ \midrule

\multicolumn{6}{c}{\textbf{Hindustani Classical Music}} \\ \midrule

\multicolumn{6}{c}{Recall Queries} \\ \midrule
\textbf{Model} & \textbf{OA} & \textbf{Inst.} & \textbf{MC} & \textbf{RC} & \textbf{CR} \\ \midrule
\textbf{MusicGen Baseline} & 1500 & 1508 & 1518 & 1523 & 1514 \\
\textbf{Mustango Baseline} & 1404 & 1393 & 1362 & 1404 & 1413 \\
\textbf{MusicGen Finetuned} & 1466 & 1440 & 1453 & 1425 & 1426 \\
\textbf{Mustango Finetuned} & 1630 & 1659 & 1668 & 1648 & 1648 \\ \midrule

\multicolumn{6}{c}{Analysis Queries} \\ \midrule
\textbf{Model} & \textbf{OA} & \textbf{Inst.} & \textbf{MC} & \textbf{RC} & \textbf{CR} \\ \midrule
\textbf{MusicGen Baseline} & 1536 & 1535 & 1556 & 1536 & 1505 \\
\textbf{Mustango Baseline} & 1467 & 1435 & 1412 & 1438 & 1480 \\
\textbf{MusicGen Finetuned} & 1426 & 1462 & 1411 & 1454 & 1452 \\
\textbf{Mustango Finetuned} & 1571 & 1566 & 1620 & 1572 & 1562 \\ \midrule

\multicolumn{6}{c}{Creativity Queries} \\ \midrule
\textbf{Model} & \textbf{OA} & \textbf{Inst.} & \textbf{MC} & \textbf{RC} & \textbf{CR} \\ \midrule
\textbf{MusicGen Baseline} & 1514 & 1508 & 1526 & 1555 & 1583 \\
\textbf{Mustango Baseline} & 1553 & 1600 & 1568 & 1561 & 1565 \\
\textbf{MusicGen Finetuned} & 1494 & 1478 & 1446 & 1456 & 1471 \\
\textbf{Mustango Finetuned} & 1439 & 1414 & 1460 & 1428 & 1381 \\ \midrule

\multicolumn{6}{c}{\textbf{Turkish Makam}} \\ \midrule

\multicolumn{6}{c}{Recall Queries} \\ \midrule
\textbf{Model} & \textbf{OA} & \textbf{Inst.} & \textbf{MC} & \textbf{RC} & \textbf{CR} \\ \midrule
\textbf{MusicGen Baseline} & 1512 & 1529 & 1556 & 1577 & 1578 \\
\textbf{Mustango Baseline} & 1531 & 1512 & 1533 & 1539 & 1511 \\
\textbf{MusicGen Finetuned} & 1530 & 1542 & 1524 & 1549 & 1543 \\
\textbf{Mustango Finetuned} & 1428 & 1417 & 1387 & 1334 & 1368 \\ \midrule

\multicolumn{6}{c}{Analysis Queries} \\ \midrule
\textbf{Model} & \textbf{OA} & \textbf{Inst.} & \textbf{MC} & \textbf{RC} & \textbf{CR} \\ \midrule
\textbf{MusicGen Baseline} & 1540 & 1527 & 1561 & 1553 & 1525 \\
\textbf{Mustango Baseline} & 1564 & 1559 & 1566 & 1597 & 1607 \\
\textbf{MusicGen Finetuned} & 1461 & 1471 & 1425 & 1433 & 1469 \\
\textbf{Mustango Finetuned} & 1436 & 1442 & 1448 & 1469 & 1399 \\ \midrule

\multicolumn{6}{c}{Creativity Queries} \\ \midrule
\textbf{Model} & \textbf{OA} & \textbf{Inst.} & \textbf{MC} & \textbf{RC} & \textbf{CR} \\ \midrule
\textbf{MusicGen Baseline} & 1508 & 1528 & 1544 & 1579 & 1604 \\
\textbf{Mustango Baseline} & 1566 & 1576 & 1578 & 1573 & 1570 \\
\textbf{MusicGen Finetuned} & 1525 & 1528 & 1512 & 1507 & 1534 \\
\textbf{Mustango Finetuned} & 1402 & 1369 & 1366 & 1342 & 1291 \\
\bottomrule
\end{tabular}}
\caption{Model Evaluation Metrics across Recall, Analysis, and Creativity Queries for Hindustani Classical Music and Turkish Makam. ELO ratings (human evaluation) have higher values as better (↑).}
\label{tab:overall_model_metrics_query}
\end{table*}

The human evaluation results in Table~\ref{tab:overall_model_metrics} and Table~\ref{tab:overall_model_metrics_query}, measured using ELO ratings, also reveal intriguing patterns across genres, models, and query types. Comparing the two genres, we observe distinct performance profiles for each model. In Hindustani Classical Music, the Mustango finetuned model emerges as the clear leader (OA: 1577), outperforming other models across most categories, particularly excelling in Melodic Contour (MC: 1623). This suggests a strong grasp of the melodic structures specific to Hindustani music. Conversely, for Turkish Makam, the MusicGen finetuned model takes the lead (OA: 1597), with both MusicGen and Mustango baseline models also performing well.


The MusicGen Baseline shows remarkable consistency across both genres, often scoring above 1500 in various categories. This suggests a robust general understanding of musical elements that transcends genre boundaries. The Mustango Baseline, while competitive, generally scores lower than MusicGen Baseline, especially in Hindustani Classical Music.

Finetuning yields mixed results across the two models. For Mustango, it significantly improves performance in Hindustani Classical but drastically reduces its effectiveness in Turkish Makam. Conversely, MusicGen's finetuning slightly lowers its performance in Hindustani Classical but enhances it in Turkish Makam. This divergence underscores the complexity of adapting models to specific musical traditions without losing generalizability.

Examining performance across query types reveals further insights. In Recall queries for Hindustani Classical, Mustango finetuned significantly outperforms other models (OA: 1630), particularly in Melody (1668). For Turkish Makam, MusicGen Baseline leads in Recall queries (OA: 1512), with MusicGen finetuned close behind (OA: 1530). This suggests that finetuning can enhance a model's ability to accurately reproduce genre-specific musical elements.

Creativity queries yield particularly interesting results, with baseline models outperforming their finetuned counterparts in both genres. In Hindustani Classical, Mustango Baseline leads (OA: 1553), while in Turkish Makam, it shares the top position with MusicGen Baseline (OA: 1508 and 1566 respectively). This suggests that finetuning, while beneficial for recall and analysis, might constrain the model's creative capabilities.

\section{Kappa Score Computation} \label{appendix:kappa_score_compute}
\begin{table}[h!]
    \centering
    \[
    \begin{array}{c|ccccc}
        & \text{A$\gg$B} & \text{A$>$B} & \text{A$=$B} & \text{A$<$B} & \text{A$\ll$B} \\ \hline
    \text{A$\gg$B} & 1 & 1 & 0.33 & 0 & 0 \\
    \text{A$>$B} & 1 & 1 & 0.67 & 0.33 & 0 \\
    \text{A$=$B} & 0.33 & 0.67 & 1 & 0.67 & 0.33 \\
    \text{A$<$B} & 0 & 0.33 & 0.67 & 1 & 1 \\
    \text{A$\ll$B} & 0 & 0 & 0.33 & 1 & 1 \\
    \end{array}
    \]
    \caption{Matrix representation of distance-based agreement score for Inter Annotator Agreement. Column represents Annotator-1's preference and Row represents Annotator-2's preference.}
    \label{tab:iaa_distance}
\end{table}
\subsection{Distance-based Computation Matrix}
The \textbf{distance-based Kappa} quantifies the absolute differences in annotations by both the annotators. Each option(except None \& NA) is assigned a value between 2 to -2 in order; A $\gg$ B, A $>$ B, A = B, A $<$ B, A $\ll$ B. After assigning values we calculate absolute distances between annotator preferences while excluding all cases which are annotated None or NA by annotators. The distance values are clipped to a maximum of 3, with agreement computed as follows : 
\begin{equation*}
    p_o^{i} = \frac{3 - d}{3}
\end{equation*}

Table \ref{tab:iaa_distance} represents the annotator preferences and the agreement score for each combination.

\subsection{Direction-based Computation Matrix}

\begin{table}[h!]
    \centering
    \[
    \begin{array}{c|ccccc}
        & \text{A$\gg$B} & \text{A$>$B} & \text{A$=$B} & \text{A$<$B} & \text{A$\ll$B} \\ \hline
    \text{A$\gg$B} & 1 & 1 & 1 & 0 & 0 \\
    \text{A$>$B} & 1 & 1 & 1 & 0 & 0 \\
    \text{A$=$B} & 1 & 1 & 1 & 1 & 1 \\
    \text{A$<$B} & 0 & 0 & 1 & 1 & 1 \\
    \text{A$\ll$B} & 0 & 0 & 1 & 1 & 1 \\
    \end{array}
    \]
    \caption{Matrix representation of direction-based agreement score for Inter Annotator Agreement. Column represents Annotator-1's preference and Row represents Annotator-2's preference.}
    \label{tab:iaa_direction}
\end{table}

The \textbf{direction-based Kappa} assesses consistency in preference order rather than the extent of preference. A disagreement is defined as only when the preference orders are reversed between the two annotators (i.e., when one annotator chooses A$<$B or A$\ll$B and the other annotator chooses B$<$A or B$\ll$A) Agreement is calculated as follows : 
\begin{equation*}
    p_o^{i} = 1 - d
\end{equation*}

Table \ref{tab:iaa_direction} shows the agreement scores between different annotations.

Observed agreement is averaged per criterion, with expected agreement(\(p_e\)) and Kappa(\(\kappa\)) calculated as follows:
\begin{equation*}
\kappa = \frac{\frac{1}{n} \sum_{i=1}^{n} p_o^i - p_e}{1 - p_e}
\end{equation*}

\section{Dataset Details} \label{appendix:dataset_details}
For Hindustani Classical, the dataset includes five instrument types—sarangi, harmonium, tabla, violin, and tanpura—along with voice. It comprises 41 ragas across two laya types: Madhya and Vilambit.

For Turkish Makam, the dataset features 15 makam-specific instruments, including the oud, tanbur, ney, davul, clarinet, kös, kudüm, yaylı tanbur, tef, kanun, zurna, bendir, darbuka, classical kemençe, rebab, and çevgen. It encompasses 93 different makams and 63 distinct usuls.

\section{ELO Ratings Computation} \label{appendix:elo}
For phase I, the total number of evaluations are 36 by each annotator and we consider each annotation as a single match. In Phase II, 63 additional annotations are conducted making a total of 135 matches for computing the ELO ratings. For every match the new rating : 
\begin{equation*}
R_i = R_i + K * (S_i - E_i)
\end{equation*}
$R_i$: Player's current Elo rating.\newline
$K$: Weighting factor that determines how much a single game affects the rating. \newline
$S_i$: Outcome of the game for the player: 1 for a win, 0.5 for a draw, and 0 for a loss.

\begin{equation*}
    E_i = \frac{1}{(1 + 10^{\frac{(R_j - R_i)}{400}})}
    , E_j = \frac{1}{(1 + 10^{\frac{(R_i - R_j)}{400}})}
\end{equation*}
$R_i$: Player- 1 current Elo rating.\newline
$R_j$: Player- 2 current Elo rating.\newline
$E_i$: Expected Elo rating of Player-1.

We use a K value of 15 for calculations due to the limited number of matches; a higher K would disproportionately weight each match and skew the ELO ratings.

\section{Annotation Tool} \label{appendix:annotation_tool}

\begin{figure*}
    \centering
    \begin{minipage}{\textwidth}
        \centering
        \includegraphics[width=1\linewidth]{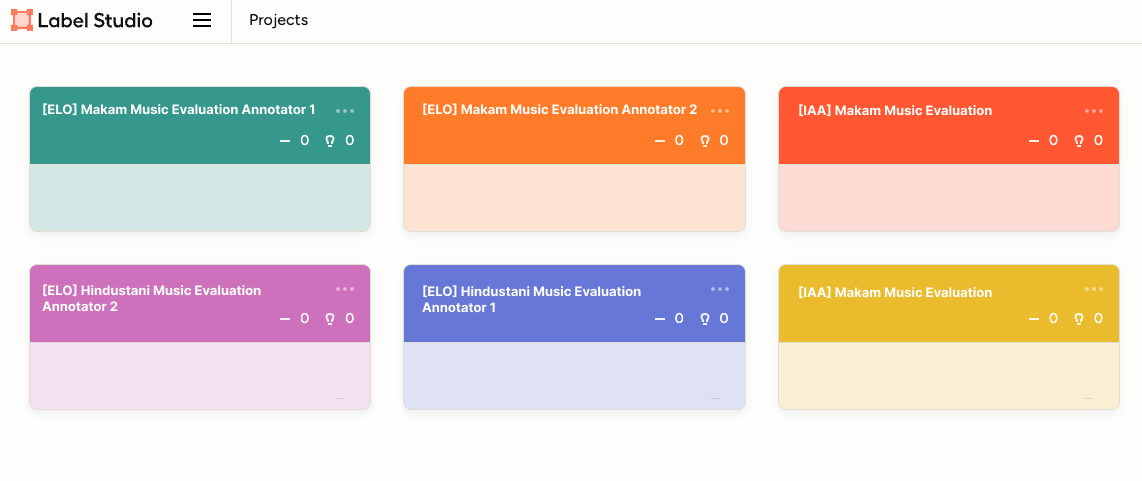}
        \label{fig:enter-label-1}
    \end{minipage}
    \vspace{1em}  
    \begin{minipage}{\textwidth}
        \centering
        \includegraphics[width=1\linewidth]{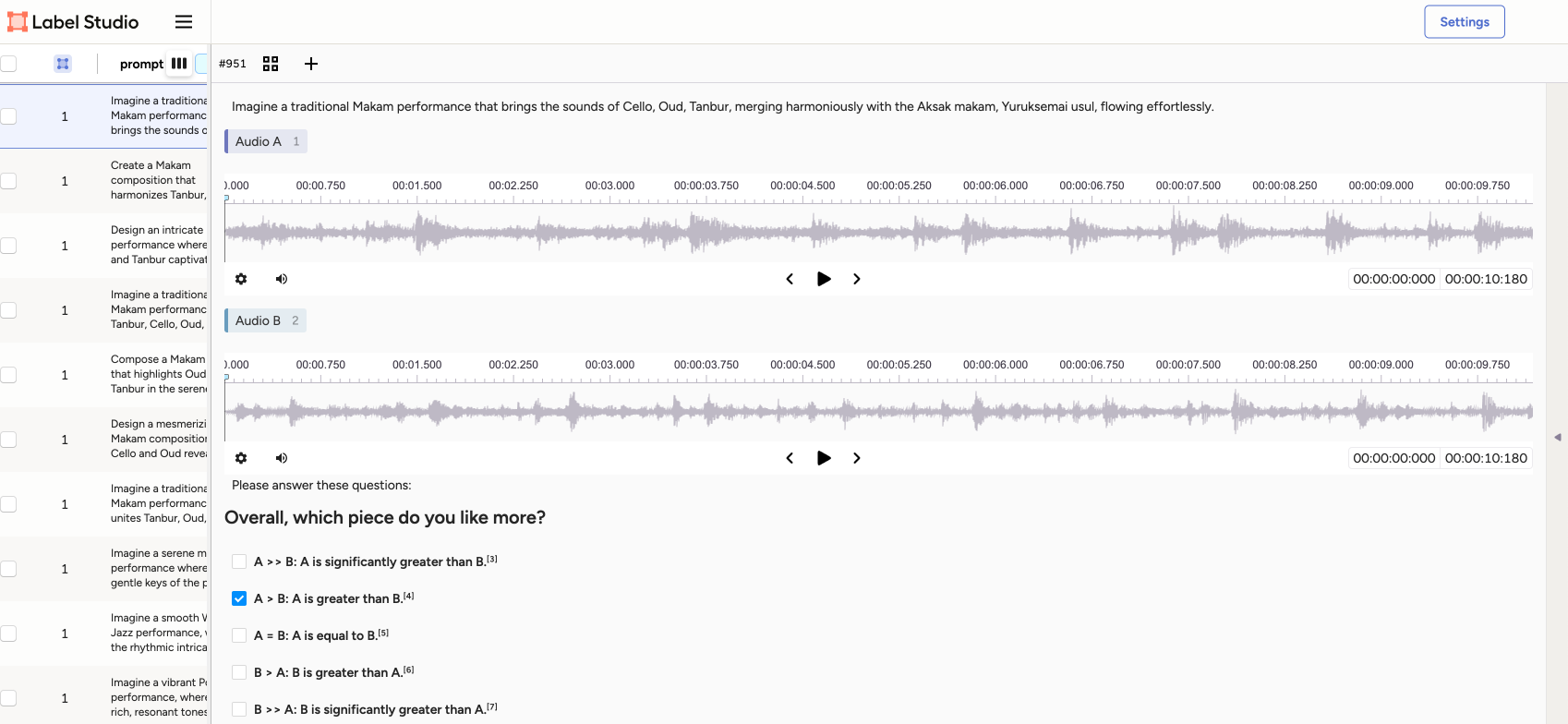}
        \caption{Screenshots of Label Studio, annotation tool for Inter Annotator Agreement and ELO rating comparison}
        \label{fig:annotation_tool}
    \end{minipage}
\end{figure*}

We deployed LabelStudio, a versatile and user-friendly annotation tool, on HuggingFace Spaces. Figure~\ref{fig:annotation_tool} provides a visual representation of our annotation tool interface, illustrating the layout and features that our human evaluators used to assess the generated music samples.

\section{Annotator Demographics} 
\label{appendix:annotator_demographics}
The annotators for our music generation task using adapter models include three individuals from India and one from Uzbekistan, all of whom are avid music listeners with diverse cultural backgrounds and a keen interest in music technology.

\end{document}